\documentclass{article}
\usepackage{spconf}
\usepackage{amsmath,amssymb,graphicx, cite, multirow}
\usepackage[table]{xcolor}
\usepackage{placeins}
\definecolor{g}{gray}{0.75}
\definecolor{lg}{gray}{0.85}

\title{Multi-MotifGAN (MMGAN): Motif-targeted Graph Generation and Prediction}
\name{Anuththari Gamage \qquad Eli Chien \qquad Jianhao Peng \qquad Olgica Milenkovic\thanks{The work was supported by the NSF Center for Science of Information under grant number 0939370}}
\address{Department of Electrical and Computer Engineering, University of Illinois Urbana-Champaign}
\begin{document}
\ninept
\maketitle
\begin{abstract}
Generative graph models create instances of graphs that mimic the properties of real-world networks. Generative models are successful at retaining pairwise associations in the underlying networks but often fail to capture higher-order connectivity patterns known as network motifs. Different types of graphs contain different network motifs, an example of which are triangles that often arise in social and biological networks. It is hence vital to capture these higher-order structures to simulate real-world networks accurately. We propose Multi-MotifGAN (MMGAN), a motif-targeted Generative Adversarial Network (GAN) that generalizes the benchmark NetGAN approach. The generalization consists of combining multiple biased random walks, each of which captures a different motif structure. MMGAN outperforms NetGAN at creating new graphs that accurately reflect the network motif statistics of input graphs such as Citeseer, Cora and Facebook. 
\end{abstract}

\section{Introduction}
\label{sec:intro}

Given the ubiquity of network structures in real-world data, graph generative models have been studied extensively as a means of simulating graphs with different properties. Classical stochastic models, such as the Erd\H{o}s-R\'enyi, Barabasi-Albert, and the stochastic block model generate graphs based on a predefined set of parameters, such as the probability of edge formation within and between communities \cite{Easley2010}. In contrast, modern approaches to graph generation based on deep learning, including NetGAN\cite{Bojchevski2018}, GraphGAN\cite{Wang2018}, and GraphRNN\cite{You2018}, are flexible enough to learn multiple different properties of an input graph simultaneously. The graphs generated by these architectures may be used for downstream learning tasks such as data augmentation\cite{chakrabarti2006}, recommendation\cite{yu2014}, and link prediction\cite{gao2011}.

Many real-world networks consist of entities with complex mutual interrelations. Such networks cannot be modeled effectively as graphs with simple pairwise relations, despite the fact that pairwise relations provide a wealth of information for learning. Studying higher-order relationships in a graph is fundamental for our understanding of the network behavior and function. Higher-order relationships are usually termed hyperedges (collections of more than two nodes) \cite{Zhou2006, chien2019} or  network motifs (recurrent node connectivity patterns that are statistically significant compared to some ground truth random graph model) \cite{Milo2002}. These higher-order structures are the actual building blocks of complex networks, as they capture fundamental functional properties.

Network motifs were originally studied in the context of gene regulatory networks \cite{Milo2002,Shen-Orr2002}, but the presence of distinct network motifs in different types of real-world networks (food webs, the world wide web, social networks, power grid networks etc.) has been established in prior literature \cite{Milo2002, Ugander2013, Dey2017}. For example, gene regulatory networks, neuronal networks, and social networks all contain a large number of triangles \cite{Milo2002, Ugander2013}. When generating graphs that are statistically similar to a real-world network or trying to predict unobserved subgraphs, it is vital to preserve the motif structures present in the network under consideration.

Existing implicit graph generative models successfully capture pairwise relationships within the graph and associated graph statistics, but they are not as successful in retaining higher-order relationships like motifs or hyperedges. To address this issue, we propose \textbf{Multi-MotifGAN (MMGAN)}, a novel motif-targeted graph generative model that preserves network motif statistics in the output graphs. MMGAN generalizes NetGAN, an architecture that uses random walks on an input graph to learn characteristics of the network. The generalization consists of combining multiple random walk statistics, where each type of random walk is biased towards one type of motif structure. We consider two variants of MMGAN: the first is designed to  reflect the motif statistics of the input graph accurately, and the second aims to improve motif prediction in networks with missing edges. Both variants combine multiple random walk outputs generated by differently biased GANs, each of which targets a specific motif type.

We show experimentally that MMGAN outperforms benchmark generative models such as NetGAN at retaining mutltiple network motif statistics  of the original graph, as evidenced by its competitive results in generation and link prediction on real-world social networks such as Citeseer, Cora, and Facebook~\cite{mccallum2000,sen2008,leskovec2012learning}. For example,when trained on Citeseer, which contains 1084 triangles, MMGAN produces networks with an  average of 1285 triangles, compared to an average of 625 produced by NetGAN. Similarly, in terms of motif prediction, MMGAN obtains an average precision of 99.29\% on Cora while NetGAN achieves 92.23\%.
For simplicity and due to space constraints, we only discuss results on motifs with up to $3$ nodes. However, it is straightforward to adapt MMGAN for another constant number of nodes.

\textbf{Relation to Existing Work:}
MMGAN uses multiple techniques for learning on graphs and combines them into a motif-aware model. Random walks on graphs  are widely used to learn the local and global topology of a graph \cite{Perozzi2014, Grover2016, Li2019}, while biased random walks
are used to characterize higher-order network structures like hyperedges and network motifs \cite{Lee2011,Tsourakakis2017, Backstrom2011, Han2016, Dayeh2012, Zhou2006}. Generative Adversarial Networks (GAN) are highly effective at learning implicit features of a data set and using these to generate realistic data samples. They are therefore a natural choice for both prediction tasks on incomplete data and sample generation. Combining GANs that provide multiple views of the same system is a new feature of our architecture and it is expected to improve the quality of inference tasks on the underlying data. There exists many methods for link prediction in networks~\cite{liben2007link}, but to the best of our knowledge, MMGAN is the only GAN-based generative and predictive model for motifs.

The paper is organized as follows: Section 2 introduces the MMGAN architecture, while Section 3 presents a summary of our experimental findings.

\vspace{-0.1in}
\section{Multi-MotifGAN}
\label{sec:method}

Let $\mathcal{G} = (\mathcal{V}, \mathcal{E})$ be a graph with node set $\mathcal{V} = \{1, \ldots, n\}$ and edge set $\mathcal{E} = \{(i,j) \mid i,j\in V \}$.  A subgraph of $\mathcal{G}$ is a graph $\mathcal{G'} =(\mathcal{V'}, \mathcal{E'})$ contained within $\mathcal{G}$ such that $\mathcal{V'} \subset \mathcal{V}$ and $\mathcal{E'}\subset \mathcal{E} \cap (\mathcal{V'} \times \mathcal{V'})$. The frequency of a subgraph $\mathcal{G'}$ is the number of appearances of subgraphs in $\mathcal{G}$ that are isomorphic to $\mathcal{G'}$. Furthermore, let $\mathcal{G}_r$ be a random graph model with the same number $n$ of nodes and the same node degree distribution as $\mathcal{G}$.  A network motif is defined as a subgraph that recurs in a network with a higher frequency than in the chosen random graph model $\mathcal{G}_r$~\cite{Milo2002}. 

\begin{figure}
    \centering
    \includegraphics[width=0.35\textwidth]{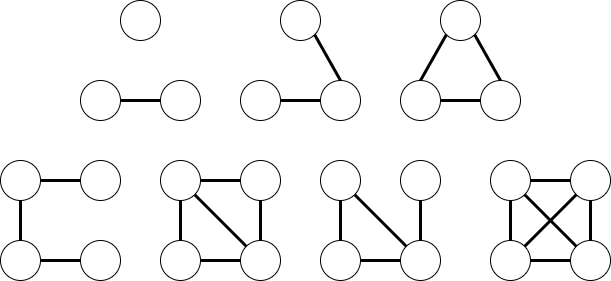}
    \caption{\emph{Motifs in social networks \cite{Ugander2013}. (Row 1) We focus on motifs with $\leq 3$ nodes: $E$ (edges), $V$ (pairs of edges sharing a vertex) and $T$ (triangles). (Row 2) Motifs involving $4$ vertices.}}
\vspace{-0.15in}
    \label{fig:motifs}
\end{figure}

\subsection{Graph Generation using NetGAN}

We base our motif-targeted generative model on an existing implicit graph generative architecture, NetGAN~\cite{Bojchevski2018}. NetGAN is a Generative Adversarial Network that uses random walks on a graph to generate realistic graphs that are statistically similar to a training graph. NetGAN consists of a generator ${G}$ and discriminator ${D}$ which are trained under the Wasserstein GAN objective~\cite{Arjovsky2017} for increased stability. The generator ${G}$ outputs sets of random walks that are similar to those sampled from an input graph that one wants to generate, while ${D}$ learns to distinguish between random walks generated by ${G}$ and those sampled from the input graph. Thus, NetGAN requires only one undirected graph as an input, from which it samples a set of random walks to act as a training data set. It is highly efficient for cases where one does not have a large set of similar graphs that can serve as the training set.

Once $G$ and $D$ are trained, NetGAN generates a new graph using the frequency of edges in the generated set of random walks. It constructs a score matrix $S$ whose $(i,j)$th entry represents the number of times edge $(i,j)$ appears in the generated random walks. The score matrix is normalized by the row sums so that for every node, one obtains a probability distribution over its neighboring nodes. To add an edge, a node is selected randomly and its neighbor is sampled according to the corresponding probability distribution constructed from the normalized score matrix. Subsequently, an edge between these two nodes is added in the output graph. The procedure continues until reaching the number of edges in the input graph.

\begin{figure}[h]
    \centering
    \includegraphics[width=0.45\textwidth]{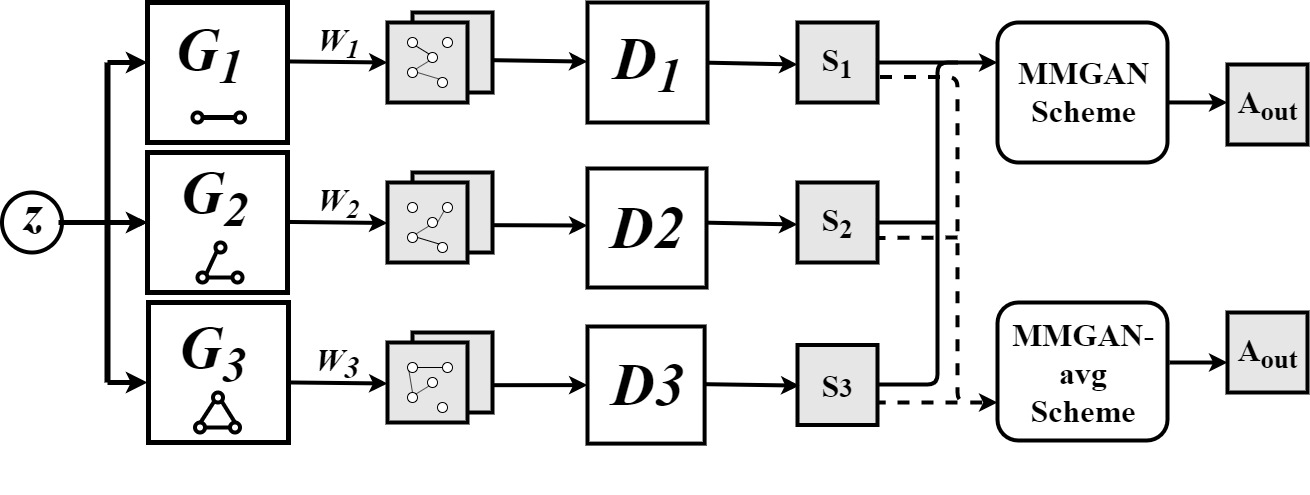}
    \caption{\emph{The MMGAN architecture, consisting of NetGAN ($G_1, D_1$), and the two motif-biased GANs ($G_2$, $D_2$) and ($G_3$, $D_3$). Each $G_i$ produces a set of random walks, while each $D_i$ determines which are plausibly coming from the input graph and generates a score matrix. The score matrices are combined under two different schemes to obtain the output graph.}}
    \label{fig:netgan}
    \vspace{-0.15in}
\end{figure}

NetGAN has been shown to outperform state-of-the-art graph generative models at preserving various topological features of the input graph (e.g. maximum degree, clustering coefficient, inter- and intra-community edge density) in its generated output. The method also exhibits competitive performance at link prediction on incomplete graphs, which indicates that it is capable of generalization rather than only memorizing the input graph. Despite the efficacy of NetGAN in the above-mentioned tasks, we observe that the graphs generated by NetGAN (as well as other state-of-the-art generative models comparable to NetGAN) fail to approximate the network motif statistics of the input graph. For example, NetGAN systematically underestimates the number of triangles in social networks by 40-60\% (see Table 1). This is a major shortcoming for applications that aim to generate graphs that realistically mimic real-world networks or predict unobserved motif structures.

\subsection{Multi-MotifGAN (MMGAN)}
For our proposed algorithm, we generalize the NetGAN random walk-based architecture which lends itself to characterizing the local properties of nodes (depending on how the random walk is performed). To generate the training set of random walks, NetGAN employs a second-order random walk, node2vec, which captures the local and global structure of the graph effectively via a two-step weighting scheme \cite{Grover2016}: given an edge $(v,x)$, suppose that the previous transition of the random walk was from some node $t$ to $v$. The second order bias $\alpha$ is chosen as
 \[\alpha_{pq}(t, x) = \begin{cases} \frac{1}{p} &   \text{ if } d_{tx} = 0,\\
1 &\text{ if } d_{tx} = 1,\\
\frac{1}{q} &
\text{ if } d_{tx} = 2,
\end{cases} \qquad\]
where $p,q \in \mathbb{R}$ and $d_{tx}$ is the shortest-length path between $t$ and $x$. The unnormalized transition probability from $v$ to $x$ equals
\[ \pi_{vx} = \alpha_{pq}(t, x) \cdot w_{vx},\]
where $w_{vx}$ is the weight of edge $(v,x)$ (equal to $1$ for unweighted graphs). In MMGAN, we change this weight to incorporate the 3-node motif statistics of the graph and bias the random walk towards edges that are more likely to be part of a particular network motif. This bias is different from the bias $\alpha_{pq}$ introduced to control the extent of exploration in the graph.

To find the correct biases, we first count the motifs in the graph of interest. While a complete enumeration of the motifs present in large-scale network is computationally prohibitive, a number of efficient motif-sampling algorithms exist that approximate the frequencies of different motif in a network \cite{Masoudi-Nejad2012,Wernicke2006}. In our analysis, we use \emph{FANMOD}, a fast network motif detection algorithm that can handle both directed and undirected networks and finds motifs containing up to $8$ nodes~\cite{Wernicke2006}. For simplicity, we focus on 3-node motifs since they represent the most significant structures in social networks and are likely to be contained in other higher-order motifs; this allows one to implicitly include information about higher-order interactions while limiting the complexity of the MMGAN platform. However, it is straightforward to adapt MMGAN to account for motifs of larger sizes with adjustments in the weight calculation and graph combination procedures so as to account for different nested non-isomorphic motifs.

Using \emph{FANMOD}, we first estimate the total number $M$ of 3-node motifs which are of types $V$ and $T$ listed in Figure~\ref{fig:motifs}. The \emph{concentration} of a motif equals $C^{(X)} = M^{(X)}/M$, where $X \in \{{V,T\}}$ and $M^{(X)}$ denotes the number of motifs of type $X$ in the graph. For an
edge $(i,j)$, we define $N^{(X)}_{ij}$
 to be the number of motifs of type $X$ in which the edge participates. Then, the \emph{motif-biased weight} of the edge $(i,j)$ equals
\[w_{ij} = \frac{\beta N^{(V)}_{ij} + (1-\beta)N^{(T)}_{ij}}{N^{(V)}_{ij} + N^{(T)}_{ij}},\]
where
$ \beta = \begin{cases}
\max\left\{C^{(V)}, C^{(T)}\right\} & (\text{ biasing towards } V)   \\
1 - \max\left\{C^{(V)}, C^{(T)}\right\} & (\text{ biasing towards } T).\\
\end{cases}$


Thus, $w_{ij}$ is a weighted average of the motif counts, weighted by an appropriate function of concentration. The chosen bias will lead to a higher frequency of the particular motif in the output graph compared to the input graph. In order to obtain motif counts that reflect the counts in the input, we combine the output score matrices of three GANs with random walks biased as follows: without using motif weights as in NetGAN ($S_1$), using weights that bias towards $V$ ($S_2$), and using weights biased towards $T$ ($S_3$). The matrix $S_1$ leads to a good characterization of the input edge set. From $S_2$, we obtain a better characterization of the $V$ motifs in the graphs when compared to $S_1$, but with a frequency that is higher. Similarly, $S_3$ ensures a good characterization of triangles, albeit once again with a higher count than observed in the input. These three `views' of the graph provide a close approximation of the actual motif frequencies and concentrations once properly combined. To handle different tasks such as motif generation and link prediction, we propose two different ways of combining the score matrices:

\textbf{I. Multi-view combination for link prediction (MMGAN-Avg):}  We combine the three score matrices via averaging, resulting in $S = \frac{S_1 + S_2+S_3}{3}$. Edges are sampled in the same manner as in NetGAN by first normalizing $S$ to produce a transition probability matrix, then selecting a node at random and choosing one of its neighbors according to the above distribution. We add an edge between the corresponding nodes in the output graph and continue adding edges similarly until the same number of edges as in the input graph is reached.

\textbf{II. Multi-view combination for graph generation (MMGAN):} In this scheme, we sample both edges and motifs from the three views at random and add them to the output graph directly as follows. We first randomly choose one view $S_i$ of $S_1, S_2, S_3$ with probabilities $p_1, p_2, p_3$ respectively. Then, we choose one of two sampling methods, sampling by maximum score or random sampling,  with probabilities $p_s$ and $1-p_s$ respectively, where $p_s$ is small to avoid overfitting.

If we choose sampling by maximum score, we first select the edge $e_i$ with the highest score in $S_i$. Then,  we add the corresponding subgraph structure to the output graph. In more detail, if $S_i = S_1$ we add $e_i$ to the output graph. If $S_i= S_2$, we find all possible $V$ motifs containing $e_i$ and compute the average score for each possible motif. Then, we select the motif with the highest average score and add the two edges of the motif to the output graph. Similarly, if $S_i = S_3$, we compute the average scores for all $T$ motifs (triangles) containing $e_i$ and add the three edges of the highest scoring motif to the output. After adding the edge(s), we remove the corresponding scores from $S_i$ to enforce sampling without replacement. We repeat this procedure with the next highest score in the score matrix and continue until the output graph has the same number of edges as the input.

If we choose random sampling, we first select a node  $n_i$ uniformly at random. Then, similar to the previous combination method, we randomly sample two other nodes $n_2, n_3$ with the probability distribution defined by the normalized score matrix. Finally, if $S_i=S_1$, we add edge $(n_1, n_2)$ to the output graph. If $S_i = S_2$, we add the $V$ motif with edges $\{(n_1, n_2), (n_1, n_3)\} $, and if  $S_i = S_3$, we add the $T$ triangle with all three nodes. We continue this until the output contains the same number of edges as the input.

Choosing some of the maximum scoring edges and motifs ensures that the key edges that appear repeatedly in the sampled set of random walks are included in the output graph. The repeated appearances indicate that the edge has a high weight and therefore is a part of a larger number of motifs. Every time we sample from these heavy-weighted edges, we add an entire motif to the output graph. Thus, adding a small sample of these will lead to a higher frequency of motifs in the output. Furthermore, by adjusting $p_1, p_2, p_3$, we can control the frequency of the different motif types as needed. This approach leads to a closer approximation of the motif counts in the original graph compared to MMGAN-avg, at the potential expense of link and motif prediction accuracy.

\section{Experimental Results}
\label{sec:results}

We test the performance of MMGAN and MMGAN-Avg against NetGAN, which is shown to outperform a number of other benchmark graph generative models in terms of preserving the input graph statistics \cite{Bojchevski2018}. For data, we use three real-world social networks, Cora \cite{sen2008}, Citeseer \cite{mccallum2000}, and Facebook \cite{leskovec2012learning} with the characteristics described in Table \ref{tab:datasets}. Note that in all of these networks, triangles ($T$) are statistically significant (occur with higher frequency in the real network compared to randomized networks). Thus, we are generally interested in keeping a comparable triangle count to the input network in our generated output.

In each of the experiments described, we train NetGAN, MMGAN, and MMGAN-Avg to 60\% edge overlap (one of the methods of early stopping in NetGAN) and average results over 5 runs. We use an 80-20\% training and testing split of the total 3-node motifs in the original graph.

\textbf{Motif-targeted
graph generation:} We evaluate the ability of MMGAN and MMGAN-Avg to preserve the motif structures in the graph by comparing  motif counts and motif concentrations in the output. For this, we combine the multiple score matrices using the combination schemes described in I and II. For II, we set  $p_1 = \frac{1}{6}, p_2 = \frac{1}{3}, p_3 = \frac{1}{2}$, emphasizing triangles, and $p_s = 0.25$ for every experiment. The choice of the probabilities is governed by the number of edges in the motifs being $1$, $2$ and $3$, respectively. The results for both combination schemes I and II are shown for comparison in Tables 1 and 2.

\textbf{Link and motif prediction:}
We evaluate the predictive ability of the MMGAN and MMGAN-Avg as follows. For motif prediction, we use the test set of motifs held out during training and construct an equally-sized set of test non-motifs. For link prediction, we use the corresponding edges as test edges and non-edges. 

\begin{table*}[h!]
\begin{center}
\begin{tabular}{*{9}{|c}|}
\hline
    \multirow{2}{*}{ Dataset} & \multirow{2}{*}{Motif} & \multirow{2}{*}{Input}  &\multicolumn{3}{c}{ Motif Count}&\multicolumn{3}{c|}{ Normalized Motif Count ($\pm$ error)} \\ \cline{4-9}
& & &  NetGAN & MMGAN & MMGAN-avg&  NetGAN & MMGAN & MMGAN-avg\\
 \hline
  \multirow{2}{*}{Citeseer}
  &  V & 22,763  & 18,369  &  23,280 & 17,464& 0.8069 ($-$0.1931)& \cellcolor{g}1.0227 ($+$0.0227)&0.7672 ($-$0.2328)\\
   &  T & 1084  & 632 & 1285 & 722 &0.5830 ($-$0.4170) & \cellcolor{g}1.1854 ($+$0.1854)& \cellcolor{lg}0.6661 ($-$0.3339)\\ \hline

     \multirow{2}{*}{Cora}
&  V & 47,239& 39,401 & 58,967 & 35,640&\cellcolor{g}0.8340 ($-$0.1660) &  1.2426 ($+$0.2426)&0.7546($-$0.2454) \\
    &  T & 1558 & 796   & 1819 & 1006&0.5110 ($-$0.4890)& \cellcolor{g}1.1675 ($+$0.1675)& \cellcolor{lg}0.6457($-$0.3543)\\ \hline

    \multirow{2}{*}{Facebook }
  &  V & 1,238,448 & 1,337,952  & 1,204,147& 1,329,432& 1.0803 ($+$0.0803) &\cellcolor{g} 0.9723($-$0.0277)
&\cellcolor{lg}1.0735($+$0.0735)\\
  &  T & 420,329 & 233,566 & 168,607 &236,144  &0.5557 ($-$0.4443)&0.4011($-$0.5989)& \cellcolor{g}0.5618($-$0.4382)\\\hline

\end{tabular}
\end{center}
\label{tab:motif-counts}
\caption{\emph{Raw motif counts in the generated graphs with normalization with respect to input count for better comparison. We use the dark shade to denote the best result (least error) over all methods and the light shade for any our methods that outperforms NetGAN.}}
\end{table*}

\begin{table*}[h!]
\begin{center}
\begin{tabular}{*{9}{|c}|}
\hline
     \multirow{2}{*}{Dataset} &\multirow{2}{*}{ Motif} & \multirow{2}{*}{Input}  &\multicolumn{3}{c|}{ Motif Concentration}&\multicolumn{3}{c|}{ KL Divergence} \\ \cline{4-9}
& & &  NetGAN & MMGAN & MMGAN-avg&  NetGAN & MMGAN & MMGAN-avg\\
 \hline
  \multirow{3}{*}{Citeseer}
& V & 95.45 \% & 96.68\%& 94.75\%& 96.03\%& \multirow{2}{*}{0.2764}& \cellcolor{lg}  & \cellcolor{g}\\
& T & 4.55\% &3.32\%&5.25\%&3.97\%&&\multirow{-2}{*}{ \cellcolor{lg} 0.0777}& \multirow{-2}{*}{ \cellcolor{g}0.0583}\\
\hline
  \multirow{3}{*}{Cora}
& V & 96.81\% &98.02\%&97.00\%&97.25\%&\multirow{2}{*}{0.3942}& \cellcolor{g}& \cellcolor{lg}\\
& T & 3.19\%&1.98\%&3.00\%&2.75\%&&\multirow{-2}{*}{ \cellcolor{g}0.0086}& \multirow{-2}{*} {\cellcolor{lg}{0.0474}}\\
\hline
   \multirow{3}{*}{Facebook}
& V & 74.66\%&85.14\%&87.72\%&84.92\%&\multirow{2}{*}{4.6922}&\multirow{2}{*}{7.5672}& \cellcolor{g}\\
& T & 25.34\%&14.86\%&12.28\%&15.08\%&&& \multirow{-2}{*}{ \cellcolor{g}4.4839}\\
\hline
\end{tabular}
\end{center}
\label{tab:motif-percs}
\caption{\emph{Motif distributions in generated graphs and comparison using Kullback-Leibler Divergence with respect to the input distribution.}}
\end{table*}
\FloatBarrier
\begin{table}[h]
    \centering
    \begin{tabular}{|c|c|c|c|c|c|c|}
    \hline
         Network  & $|\mathrm{N}|$ &  $|\mathrm{E}|$  & $C^{(V)}$ & $C^{(T)}$ &$R^{(V)}$ & $R^{(T)}$   \\ \hline
         Cora & 2485 & 10,138 & 96.81 &  3.19 & 99.97  & 0.03 \\
         \hline
         Citeseer & 2118 & 7358 & 95.45 & 4.55 & 99.94 & 0.06\\
         \hline
        Facebook & 1034 & 53,498  & 74.66 & 25.34 & 96.68 & 3.32 \\
         \hline

    \end{tabular}
    \caption{\emph{Statistics of the real-world network used for testing. $|\mathrm{N}|$ and $|\mathrm{E|}$ are the number of nodes and edges in the largest connected component of the graph respectively. $C^{(V)}$ and $C^{(T)}$ represent the concentration of each motif (proportion of motifs of each type in the total set of 3-node motifs). $R^{(V)}$ and $R^{(T)}$ show the average concentration of each motif type in a set of graphs drawn from the random graph model $\mathcal{G}_r$.}}
    \label{tab:datasets}
\end{table}

We use the average scores of these test motifs and edges to compute two metrics: AUC (Area Under the Curve of the Receiver Operating Characteristic) and AP (Average Precision), which are standard metrics for link prediction evaluation~\cite{Bojchevski2018}. Tables 4 and 5 show the results under each metric.

\section{Discussion}
While all three algorithms are quite successful, MMGAN-Avg outperforms all the other methods in every dataset under all metrics and should be the method of choice for motif prediction. The two different GAN-combining schemes essentially tradeoff between exploration and exploitation in different manners. MMGAN targets edges that are more likely to produce motifs and adds them to the output, thus ensuring that we obtain close to the input counts. MMGAN-Avg on the other hand incorporates information from all three views equally, resulting in a graph that better reflects the edge connectivity of the input network. Nevertheless, it appears plausible that large-scale tuning of the motif sampling probabilities and the proportions of the maximum and random score selection in MMGAN may lead to improved performance compared to MMGAN-Avg. These will be described in the full version of the paper. 

We further note that even without explicitly incorporating statistics of 4-node motifs in the input network, MMGAN approximates their counts better than NetGAN. For example, we compare the square (4-node cycle) counts in the output when they were trained on Citeseer. NetGAN generates graphs that have a normalized count of 0.1204 on average, while MMGAN has a normalized count of 0.3012 on average in its output graphs. This supports our assumption that since 3-node motifs are  likely to be contained in other higher-order motifs, using only the 3-node motif statistics still allows us to implicitly include information about the higher-order motifs.

\begin{table}[h!]
\begin{center}
\begin{tabular}{|c|c|c|c|c|}
\hline
  Dataset& Type & NetGAN & MMGAN & MMGAN-avg
 \\ \hline
  \multirow{2}{*}{Citeseer}
  & Link& 0.9599& 0.9265& \cellcolor{g}0.9675 \\
&  Motif & 0.9974  & 0.9958&\cellcolor{g} 0.9982\\ \hline
  \multirow{2}{*}{Cora}
 & Link &  0.9159 & 0.8947&\cellcolor{g} 0.9340\\
&  Motif &0.9961 &0.9907 &\cellcolor{g}0.9977\\ \hline
  \multirow{2}{*}{Facebook}
 & Link & 0.9779 & 0.9751 & \cellcolor{g}0.9981\\
&  Motif &0.9733
&0.9585 &\cellcolor{g}0.9770\\ \hline
\end{tabular}
\end{center}
\label{tab:AUC}
\caption{\emph{Link and motif prediction quality measured using Area Under the Curve (AUC).}}
\end{table}

\begin{table}[h!]
\begin{center}
\begin{tabular}{|c|c|c|c|c|}
\hline
  Dataset& Type & NetGAN & MMGAN & MMGAN-avg
 \\ \hline
  \multirow{2}{*}{Citeseer}
  & Link& 0.9655   & 0.9391& \cellcolor{g}0.9730\\
&  Motif & 0.9962 & 0.9950 &\cellcolor{g}0.9970\\ \hline
  \multirow{2}{*}{Cora}
 & Link & 0.9223 & 0.9010 &\cellcolor{g}0.9429 \\
&  Motif & 0.9959 & 0.9902 &\cellcolor{g}0.9969\\ \hline
  \multirow{2}{*}{Facebook}
 & Link & 0.9735 & 0.9743& \cellcolor{g}0.9816\\
&  Motif & 0.9578 & 0.9337&\cellcolor{g}0.9632\\ \hline
\end{tabular}
\end{center}
\label{tab:AP}
\caption{\emph{Link and motif prediction quality measured using Average Precision.}}
\end{table}
\vspace{-0.15in}

\bibliographystyle{IEEEbib}
\bibliography{biblio}

\begin{thebibliography}{10}

\bibitem{Easley2010}
D.~Easley and J.~Kleinberg,
\newblock {\em Networks, Crowds, and Markets: Reasoning About a Highly
  Connected World},
\newblock Cambridge University Press, New York, NY, USA, 2010.

\bibitem{Bojchevski2018}
A.~Bojchevski, O.~Shchur, D.~Z{\"u}gner, and S.~G{\"u}nnemann,
\newblock ``{N}et{GAN}: Generating graphs via random walks,''
\newblock in {\em Proceedings of the 35th International Conference on Machine
  Learning}, Jennifer Dy and Andreas Krause, Eds., Stockholmsmässan, Stockholm
  Sweden, 10--15 Jul 2018, vol.~80 of {\em Proceedings of Machine Learning
  Research}, pp. 610--619, PMLR.

\bibitem{Wang2018}
H.~Wang, J.~Wang, J.~Wang, M.~Zhao, W.~Zhang, F.~Zhang, X.~Xie, and M.~Guo,
\newblock ``Graph{GAN}: Graph representation learning with generative
  adversarial nets,''
\newblock in {\em AAAI Conference on Artificial Intelligence (AAAI-18)}, 2018.

\bibitem{You2018}
J.~You, R.~Ying, X.~Ren, W.~Hamilton, and J.~Leskovec,
\newblock ``{G}raph{RNN}: Generating realistic graphs with deep auto-regressive
  models,''
\newblock in {\em Proceedings of the 35th International Conference on Machine
  Learning}, Stockholmsmässan, Stockholm Sweden, 10--15 Jul 2018, vol.~80 of
  {\em Proceedings of Machine Learning Research}, pp. 5708--5717, PMLR.

\bibitem{chakrabarti2006}
D.~Chakrabarti and C.~Faloutsos,
\newblock ``Graph mining: Laws, generators, and algorithms,''
\newblock {\em ACM computing surveys (CSUR)}, vol. 38, no. 1, pp. 2, 2006.

\bibitem{yu2014}
X.~Yu, X.~Ren, Y.~Sun, Q.~Gu, B.~Sturt, U.~Khandelwal, B.~Norick, and J.~Han,
\newblock ``Personalized entity recommendation: A heterogeneous information
  network approach,''
\newblock in {\em Proceedings of the 7th ACM international conference on Web
  search and data mining}. ACM, 2014, pp. 283--292.

\bibitem{gao2011}
S.~Gao, L.~Denoyer, and P.~Gallinari,
\newblock ``Temporal link prediction by integrating content and structure
  information,''
\newblock in {\em Proceedings of the 20th ACM international conference on
  Information and knowledge management}. ACM, 2011, pp. 1169--1174.

\bibitem{Zhou2006}
D.~Zhou, J.~Huang, and B.~Sch\"{o}lkopf,
\newblock ``Learning with hypergraphs: Clustering, classification, and
  embedding,''
\newblock in {\em Advances in Neural Information Processing Systems 19},
  B.~Sch\"{o}lkopf, J.~C. Platt, and T.~Hoffman, Eds., pp. 1601--1608. MIT
  Press, 2007.

\bibitem{chien2019}
E.~Chien, P.~Li, and O.~Milenkovic,
\newblock ``Landing probabilities of random walks for seed-set expansion in
  hypergraphs,''
\newblock {\em CoRR}, vol. abs/1910.09040, 2019.

\bibitem{Milo2002}
R.~Milo, S.~Shen-Orr, S.~Itzkovitz, N.~Kashtan, D.~Chklovskii, and U.~Alon,
\newblock ``Network motifs: Simple building blocks of complex networks,''
\newblock {\em Science}, vol. 298, pp. 824--827, 2002.

\bibitem{Shen-Orr2002}
S.~Shen-Orr, R.~Milo, S.~Mangan, and U.~Alon,
\newblock ``Network motifs in the transcriptional regulation network of
  escherichia coli,''
\newblock {\em Nature Genetics}, vol. 31, no. 1, pp. 64--68, 2002.

\bibitem{Ugander2013}
J.~Ugander, L.~Backstrom, and J.~Kleinberg,
\newblock ``Subgraph frequencies: Mapping the empirical and extremal geography
  of large graph collections,''
\newblock in {\em Proceedings of the 22Nd International Conference on World
  Wide Web}, New York, NY, USA, 2013, pp. 1307--1318, ACM.

\bibitem{Dey2017}
A.~K. {Dey}, Y.~R. {Gel}, and H.~V. {Poor},
\newblock ``Motif-based analysis of power grid robustness under attacks,''
\newblock in {\em 2017 IEEE Global Conference on Signal and Information
  Processing (GlobalSIP)}, Nov 2017, pp. 1015--1019.

\bibitem{mccallum2000}
A.~K. McCallum, K.~Nigam, J.~Rennie, and K.~Seymore,
\newblock ``Automating the construction of internet portals with machine
  learning,''
\newblock {\em Information Retrieval}, vol. 3, pp. 127--163, 2000.

\bibitem{sen2008}
P.~Sen, G.~Namata, M.~Bilgic, L.~Getoor, B.~Galligher, and T.~Eliassi-Rad,
\newblock ``Collective classification in network data,''
\newblock {\em AI magazine}, vol. 29, no. 3, pp. 93--93, 2008.

\bibitem{leskovec2012learning}
J.~Leskovec and J.~Mcauley,
\newblock ``Learning to discover social circles in ego networks,''
\newblock in {\em Advances in neural information processing systems}, 2012, pp.
  539--547.

\bibitem{Perozzi2014}
B.~Perozzi, R.~Al-Rfou, and S.~Skiena,
\newblock ``Deepwalk: Online learning of social representations,''
\newblock in {\em Proceedings of the 20th ACM SIGKDD International Conference
  on Knowledge Discovery and Data Mining}, New York, NY, USA, 2014, KDD '14,
  pp. 701--710, ACM.

\bibitem{Grover2016}
A.~Grover and J.~Leskovec,
\newblock ``Node2vec: Scalable feature learning for networks,''
\newblock in {\em Proceedings of the 22Nd ACM SIGKDD International Conference
  on Knowledge Discovery and Data Mining}, New York, NY, USA, 2016, pp.
  855--864, ACM.

\bibitem{Li2019}
P.~Li, I~Chien, and O.~Milenkovic,
\newblock ``{Optimizing Generalized PageRank Methods for Seed-Expansion
  Community Detection},''
\newblock {\em CoRR}, vol. abs/1905.10881, 2019.

\bibitem{Lee2011}
J.~{Lee}, M.~{Cho}, and K.~M. {Lee},
\newblock ``Hyper-graph matching via reweighted random walks,''
\newblock in {\em CVPR 2011}, June 2011, pp. 1633--1640.

\bibitem{Tsourakakis2017}
C.~E. Tsourakakis, J.~Pachocki, and M.~Mitzenmacher,
\newblock ``Scalable motif-aware graph clustering,''
\newblock in {\em Proceedings of the 26th International Conference on World
  Wide Web}, Republic and Canton of Geneva, Switzerland, 2017, WWW '17, pp.
  1451--1460, International World Wide Web Conferences Steering Committee.

\bibitem{Backstrom2011}
L.~Backstrom and J.~Kleinberg,
\newblock ``Network bucket testing,''
\newblock in {\em Proceedings of the 20th International Conference on World
  Wide Web}, New York, NY, USA, 2011, pp. 615--624, ACM.

\bibitem{Han2016}
G.~{Han} and H.~{Sethu},
\newblock ``Waddling random walk: Fast and accurate mining of motif statistics
  in large graphs,''
\newblock in {\em 2016 IEEE 16th International Conference on Data Mining
  (ICDM)}, Dec 2016, pp. 181--190.

\bibitem{Dayeh2012}
M.~{El Dayeh} and M.~{Hahsler},
\newblock ``Biological pathway completion using network motifs and random walks
  on graphs,''
\newblock in {\em 2012 IEEE Symposium on Computational Intelligence in
  Bioinformatics and Computational Biology (CIBCB)}, May 2012, pp. 229--236.

\bibitem{liben2007link}
D.~Liben-Nowell and J.~Kleinberg,
\newblock ``The link-prediction problem for social networks,''
\newblock {\em Journal of the American society for information science and
  technology}, vol. 58, no. 7, pp. 1019--1031, 2007.

\bibitem{Arjovsky2017}
M.~Arjovsky, S.~Chintala, and L{\'e}on Bottou,
\newblock ``{W}asserstein generative adversarial networks,''
\newblock in {\em Proceedings of the 34th International Conference on Machine
  Learning}, Doina Precup and Yee~Whye Teh, Eds., International Convention
  Centre, Sydney, Australia, 06--11 Aug 2017, vol.~70 of {\em Proceedings of
  Machine Learning Research}, pp. 214--223, PMLR.

\bibitem{Masoudi-Nejad2012}
A.~{Masoudi-Nejad}, F.~{Schreiber}, and Z.~R.~M. {Kashani},
\newblock ``Building blocks of biological networks: a review on major network
  motif discovery algorithms,''
\newblock {\em IET Systems Biology}, vol. 6, no. 5, pp. 164--174, Oct 2012.

\bibitem{Wernicke2006}
S.~Wernicke and F.~Rasche,
\newblock ``{FANMOD}: A tool for fast network motif detection,''
\newblock {\em Bioinformatics}, vol. 22, no. 9, pp. 1152--1153, May 2006.

\end{thebibliography}

\end{document}